\documentstyle[epsfig]{aipproc}

\newcommand{\be}{\begin{equation}}
\newcommand{\ee}{\end{equation}}
\newcommand{\beqn}{\begin{eqnarray}}
\newcommand{\eeqn}{\end{eqnarray}}
\newcommand{\fP}{I\!\!P}
\newcommand{\spc}[1]{\mbox{\hspace{#1}}}

\begin{document}
\setlength{\unitlength}{1cm}

\title{Large Rapidity Gap Events in DIS}

\author{Mark W\"usthoff}
\address{High Energy Physics Division,
         Argonne National Laboratory, Argonne, IL 60439, USA}
\maketitle

\begin{abstract}
Diffractive scattering in DIS is discussed in terms of the
perturbative two-gluon model and numerical results for 
$F_2^D$ are presented.
\end{abstract}

Large Rapidity Gap Events can only be explained in conjunction with color zero
or vacuum exchange, since gaps with color exchange are exponentially 
suppressed. The vacuum exchange is generally divided up into two groups, the
subleading meson exchanges and the Pomeron which represents the leading  
exchange at very high energies or small x-Bjorken.
In most cases the proton emerges intact from the reaction
(Single Diffractive Dissociation) and proton dissociation 
(or Double Diffractive Dissociation) is only considered 
as background which is, however, not completely negligible.
\begin{figure}[h]
  \begin{center}
    \leavevmode
\begin{picture}(0,0)%
\epsfig{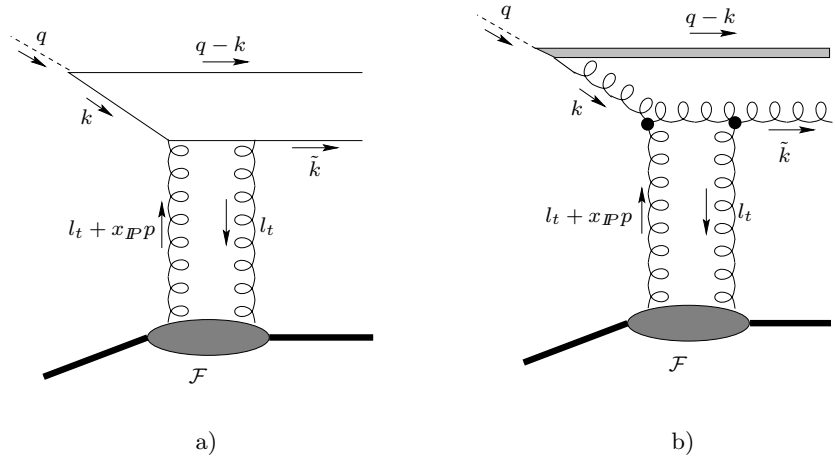}%
\end{picture}%
\setlength{\unitlength}{0.00025000in}%
\begingroup\makeatletter\ifx\SetFigFont\undefined
\def\x#1#2#3#4#5#6#7\relax{\def\x{#1#2#3#4#5#6}}%
\expandafter\x\fmtname xxxxxx\relax \def\y{splain}%
\ifx\x\y   
\gdef\SetFigFont#1#2#3{%
  \ifnum #1<17\tiny\else \ifnum #1<20\small\else
  \ifnum #1<24\normalsize\else \ifnum #1<29\large\else
  \ifnum #1<34\Large\else \ifnum #1<41\LARGE\else
     \huge\fi\fi\fi\fi\fi\fi
  \csname #3\endcsname}%
\else
\gdef\SetFigFont#1#2#3{\begingroup
  \count@#1\relax \ifnum 25<\count@\count@25\fi
  \def\x{\endgroup\@setsize\SetFigFont{#2pt}}%
  \expandafter\x
    \csname \romannumeral\the\count@ pt\expandafter\endcsname
    \csname @\romannumeral\the\count@ pt\endcsname
  \csname #3\endcsname}%
\fi
\fi\endgroup
\begin{picture}(17303,9458)(1465,-10045)
\put(13516,-2796){\makebox(0,0)[lb]{\smash{\SetFigFont{8}{9.6}{rm}$k$}}}
\put(15977,-827){\makebox(0,0)[lb]{\smash{\SetFigFont{8}{9.6}{rm}$q-k$}}}
\put(12018,-653){\makebox(0,0)[lb]{\smash{\SetFigFont{8}{9.6}{rm}$q$}}}
\put(5476,-9961){\makebox(0,0)[lb]{\smash{\SetFigFont{9}{10.8}{rm}a)}}}
\put(15676,-9961){\makebox(0,0)[lb]{\smash{\SetFigFont{9}{10.8}{rm}b)}}}
\put(7901,-4111){\makebox(0,0)[lb]{\smash{\SetFigFont{8}{9.6}{rm}$\tilde{k}$}}}
\put(17971,-3811){\makebox(0,0)[lb]{\smash{\SetFigFont{8}{9.6}{rm}$\tilde{k}$}}}
\put(15623,-8298){\makebox(0,0)[lb]{\smash{\SetFigFont{8}{9.6}{rm}$\cal{F}$}}}
\put(17104,-5013){\makebox(0,0)[lb]{\smash{\SetFigFont{8}{9.6}{rm}$l_t$}}}
\put(12981,-5086){\makebox(0,0)[lb]{\smash{\SetFigFont{8}{9.6}{rm}$l_t+x_{\fP}p$}}}
\put(5328,-8553){\makebox(0,0)[lb]{\smash{\SetFigFont{8}{9.6}{rm}$\cal{F}$}}}
\put(6853,-5312){\makebox(0,0)[lb]{\smash{\SetFigFont{8}{9.6}{rm}$l_t$}}}
\put(2751,-5386){\makebox(0,0)[lb]{\smash{\SetFigFont{8}{9.6}{rm}$l_t+x_{\fP}p$}}}
\put(3001,-2986){\makebox(0,0)[lb]{\smash{\SetFigFont{8}{9.6}{rm}$k$}}}
\put(5477,-1427){\makebox(0,0)[lb]{\smash{\SetFigFont{8}{9.6}{rm}$q-k$}}}
\put(2052,-1212){\makebox(0,0)[lb]{\smash{\SetFigFont{8}{9.6}{rm}$q$}}}
\end{picture}
  \end{center}
  \caption{Quark dipole (a) and gluon dipole (b).}
  \label{fig1}
\end{figure}

The simplest representation of the Pomeron within QCD is a two
gluon pair with zero net color charge. The full structure of the Pomeron
is of cause much more complicated and higher order contributions have
to be added. Fortunately, the first correction, which is due to 
multi gluon exchange,
does not require any recalculation as will be explained. 
The process depicted in Fig.\ref{fig1} consists of two basic  
contributions, the production of a single quark-antiquark pair and 
the emission of an extra gluon.
Higher order multi parton final states may be included by switching on 
QCD-evolution, in this talk, however, only the lower orders are
considered.

A very convenient and intuitive approach is the wave function formalism
which associates the parton content of the photon with a color dipole.
The simple quark-antiquark dipole yields the following wave function:
\beqn\label{lw1}
\Psi^{\gamma}_h(\alpha,k_t)
&=&\left\{ \begin{array}{ll}
\begin{displaystyle}
\frac{\sqrt{2}\;(\alpha -1)\;k_t}{|k_t|^2+\alpha (1-\alpha)Q^2}
\end{displaystyle}&\;\mbox{for $\gamma=+1$ and $h=+1$}
\\ & \\ \begin{displaystyle}
\frac{\sqrt{2}\; \alpha\;k_t}{|k_t|^2+\alpha (1-\alpha)Q^2}
\end{displaystyle}&\;\mbox{for $\gamma=+1$ and $h=-1$}
\\ & \\ \begin{displaystyle}
\frac{\sqrt{2}\; \alpha\;k_t^*}{|k_t|^2+\alpha (1-\alpha)Q^2}
\end{displaystyle}&\;\mbox{for $\gamma=-1$ and $h=+1$}
\\ & \\ \begin{displaystyle}
\frac{\sqrt{2}\;(\alpha -1)\;k_t^*}{|k_t|^2+\alpha (1-\alpha)Q^2}
\end{displaystyle}&\;\mbox{for $\gamma=-1$ and $h=-1$}
\end{array} \right. \\  \nonumber \; 
\eeqn
and
\beqn\label{lw2}
\Psi^{\gamma}_h(\alpha,k_t)&=& \spc{0.5cm} 2\;\frac{\alpha (1-\alpha) \;Q}
{|k_t|^2+\alpha (1-\alpha)Q^2} \spc{0.48cm} \mbox{for $\gamma=\ 0$ and 
$h=\pm 1$}
\eeqn
where $\gamma$ denotes the photon and $h$ the quark helicity.
The variables correspond to the Sudakov-decomposition of the momentum
$\tilde{k}$ (see Fig.\ref{fig1}):
$\tilde{k}={\alpha}Q'+\frac{k_t^2}{{\alpha}s}p+k_t$. 
Similar results can be found in \cite{Mue,NZ,BLW}

In the case of gluon emission one finds that for large enough
photon virtuality $Q^2$ the quark-antiquark state (the shaded area in
Fig.\ref{fig1}.b)  effectively acts as a gluon state.
This leads to the notion of a single gluon dipole which
in analogy to the quark-antiquark dipole previously 
is described by the following wave function:
\be\label{lw5}
\Psi^{mn}(\alpha,k_t)\;=\;\frac{1}{\sqrt{\alpha(1-\alpha)Q^2}}\;
\frac{k_t^2\;\delta^{mn}\;-\;2\;k_t^mk_t^n}{k_t^2+\alpha(1-\alpha)Q^2}\;\;.
\ee
This expression goes beyond earlier results in \cite{Mue,NZ,Rys} 
which were restricted to the triple Regge limit (large 
diffractive mass) where $\alpha(1-\alpha)Q^2$ 
is much smaller than $k_t^2$. In this case the $\delta^{mn}$-term
drops out when all diagrams are summed up. It, however, becomes
important when the extension towards small masses is considered.

In both cases the wave function has to be folded with the unintegrated
gluon structure function ${\cal F}$ (see Fig.\ref{fig1})
which can be extracted from the
inclusive $F_2$-data (for more details see \cite{Wu}). 
The t-dependence (momentum transfer) was added by hand using
the diffractive slope as measured by ZEUS\cite{Grothe}.
It is interesting to note that the semiclassical
approach of ref.\cite{Arthur} obtains the same results as presented
in eqs.(1-3). The identity
\be W(k_{\perp})\;=\;2\;\frac{{\cal F}(k_{\perp})}{k^2_{\perp}} 
- 2\;\delta^2(k_{\perp})
\int d^2k'_{\perp} \frac{{\cal F}(k'_{\perp})}{k'^2_{\perp}} 
\ee
provides the link between the classical gluon field $W$ in ref.\cite{Arthur} 
and the unintegrated structure function ${\cal F}$.
A perturbative expansion of the underlying eikonal factors
shows that $W$ can be interpreted as multi gluon exchange. It therefore 
represents a major improvement over the simple two-gluon model. This is 
of particular relevance, since the diffractive process is
dominated by low $k_t$, i.e. soft contributions.

In the following section the main numerical results for 
$F_2^D(x_{\fP},\beta,Q^2)$ as derived from eqs.(1-3) in combination
with an 
appropriate parametrization for the unintegrated structure function ${\cal F}$ 
are presented ($\beta=Q^2/(M^2+Q^2)$ and $x_{\fP}=(M^2+Q^2)/(W^2+Q^2)$).
\begin{figure}[h]
\begin{center}
\leavevmode
\epsfig{file=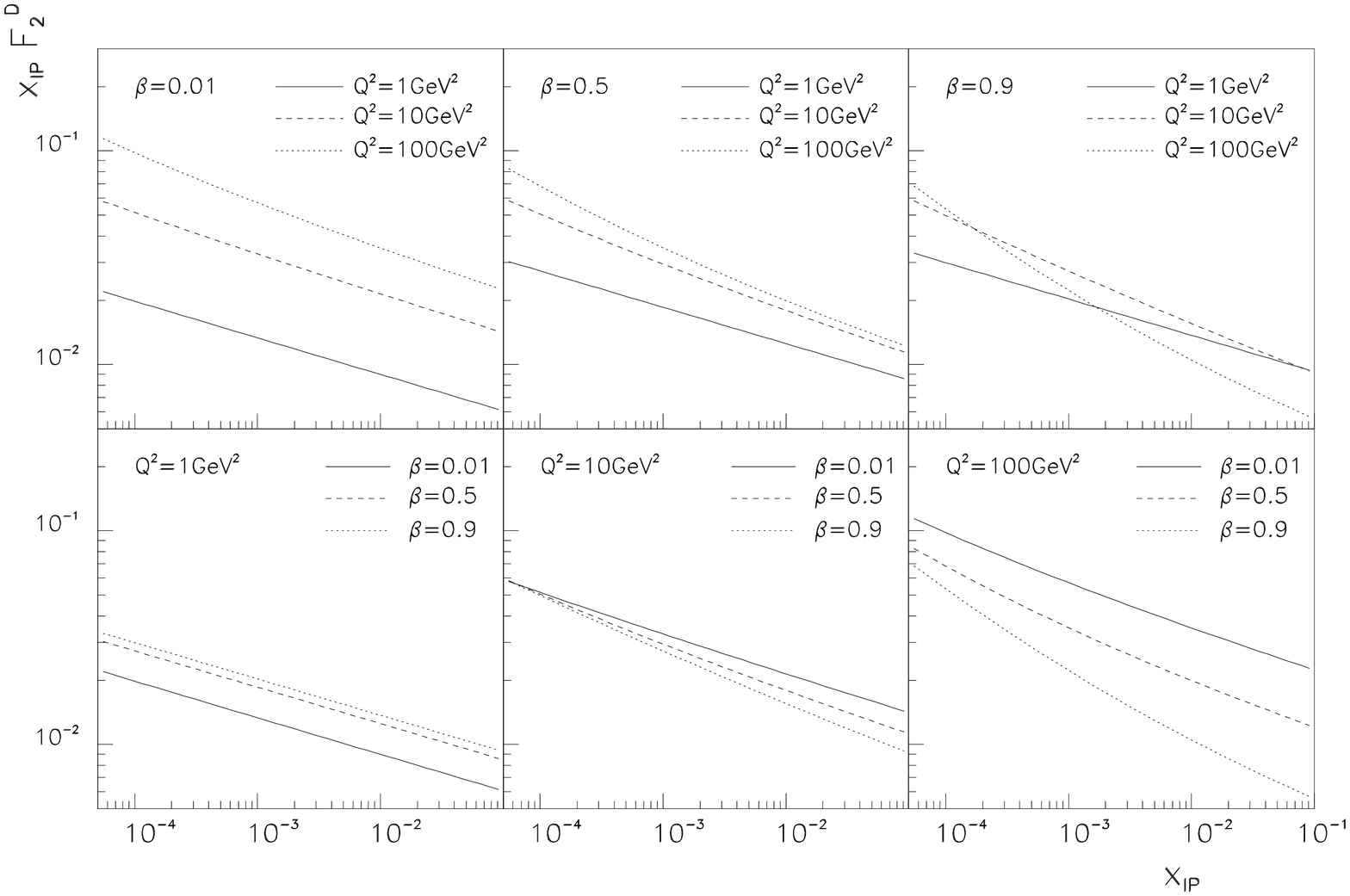,height=7cm}
\end{center}
\caption{$x_{\fP}$-distribution.}
\label{xpfig}
\end{figure}
First we examine the $x_{\fP}$-distribution
(Fig.\ref{xpfig}). The Pomeron
intercept dependends on the scale $k_t^2/(1-\beta)$
which leads to the breaking of Regge-factorization. With increasing
$Q^2$ and/or increasing $\beta$ the $x_{\fP}$-distributions become
steeper, i.e. the process harder. Since the scale is not directly
coupled to $Q^2$ the process never becomes purely hard and the
Pomeron-intercept lies only slightly above the value for the soft
Pomeron.

\begin{figure}[h]
\begin{center}
\leavevmode
\epsfig{file=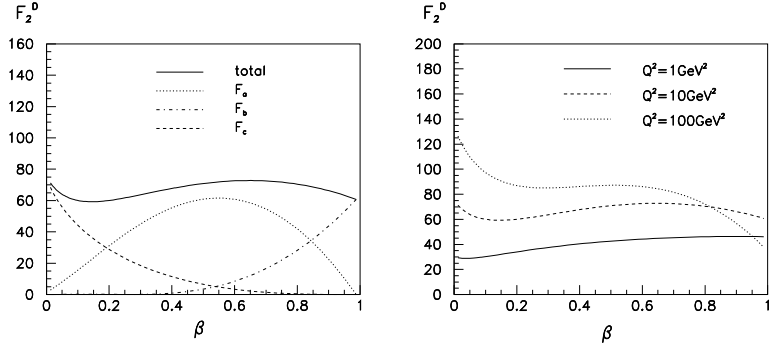,height=5cm}
\end{center}
\caption{$\beta$-spectrum at $x_{\fP}=5\cdot10^{-4}$.}
\label{betafig}
\end{figure}

The shape of the $\beta$-spectrum turns out to be rather flat when 
all three contributions associated with the longitudinal 
and transverse production of quarks (large and medium 
$\beta$, $F_b$ and $F_a$ in Fig.\ref{betafig}.a)
and the emission of an extra gluon are combined.
The latter fills the low and only the low $\beta$-region
($F_c$ in Fig.\ref{betafig}.a). The fact that this contribution is rapidly
decreasing with increasing $\beta$ is an
important consequence of the $k_t^2\delta^{mn}$-term in eq.(3).
With increasing $Q^2$ the small $\beta$ region starts rising
(log($Q^2$)-correction) whereas the large $\beta$ region decreases
which indicates the higher twist nature of the longitudinal part.

\begin{figure}[h]
\begin{center}
\begin{picture}(14,5)
\leavevmode
\put(0,0){\epsfig{file=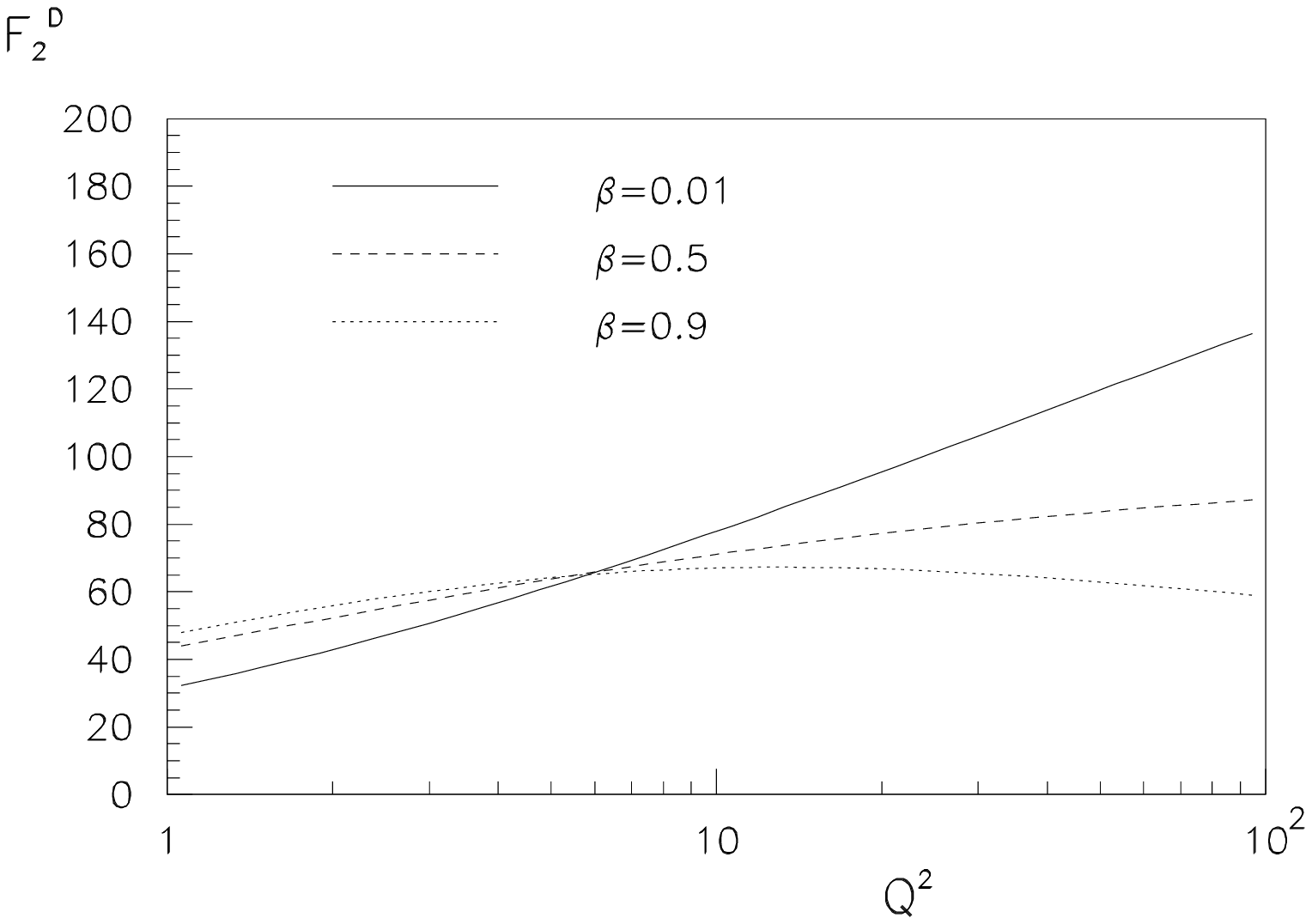,height=5cm}}
\put(7,0){\epsfig{file=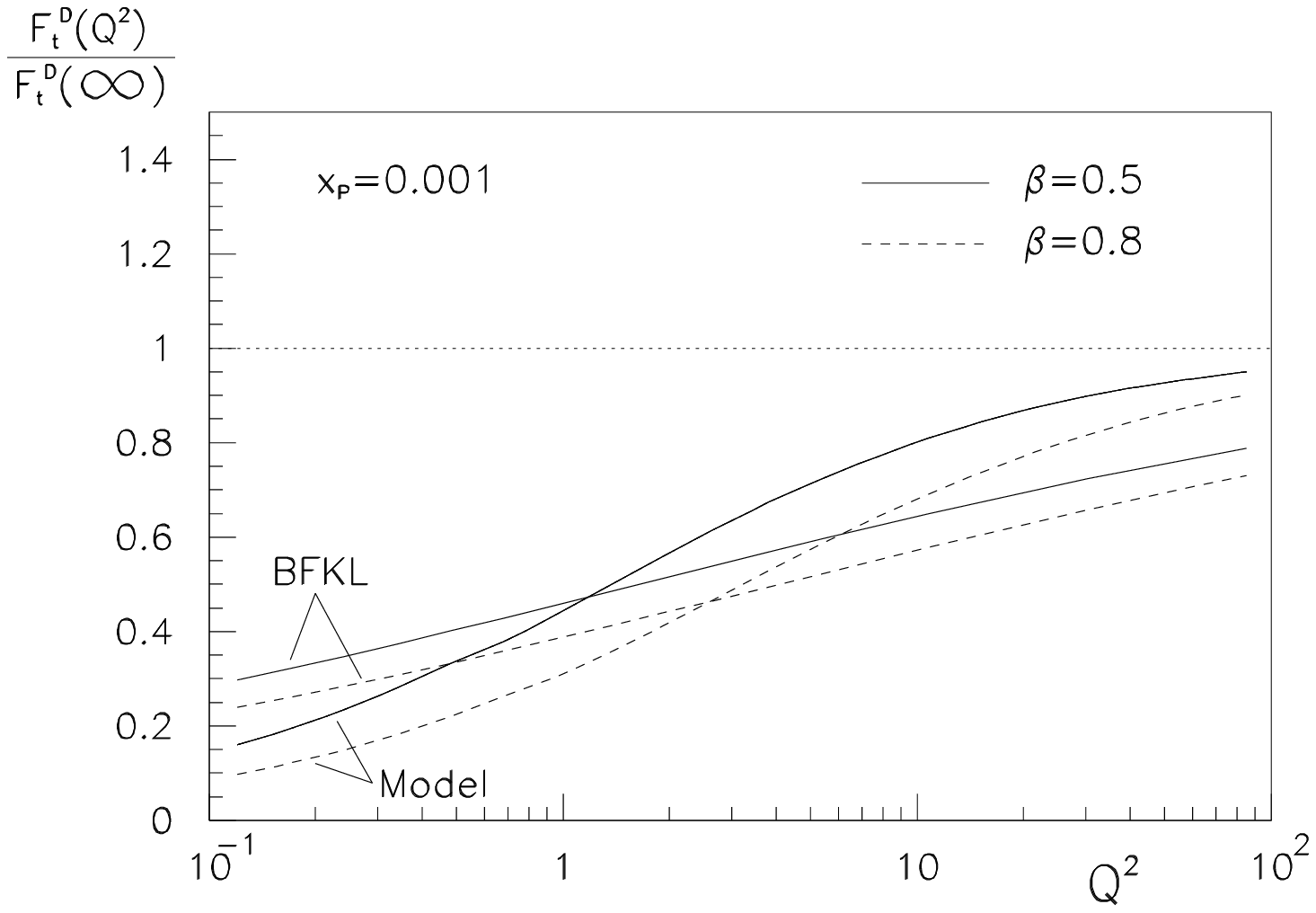,height=5cm}}
\end{picture}
\end{center}
\caption{$Q^2$-dependence.}
\label{q2fig}
\end{figure}

The $Q^2$-dependence is more clearly depicted in
Fig.\ref{q2fig}.a. Due to the $Q^2$ cutoff in the phase space all
graphs show in the beginning a rise with $Q^2$. Only at rather 
large $Q^2$ the asymptotic regime is approached and 
the higher twist contribution (dominant when $\beta=0.9$) turns down 
while the purely leading twist part (dominant when $\beta=0.5$) flattens out. 
This delay in the asymptotic behavior is model dependent as can be seen
in Fig.\ref{q2fig}.b. Only the transverse part
of the quark-antiquark production which is supposed to give a flat
leading twist contribution is presented here. 
The phenomenological Pomeron model as
compared to the BFKL-Pomeron \cite{BFKL} reaches the 
asymptotic limit faster.
The observed deviations from the leading twist behavior are in general
related to higher twist corrections which turn out be non negligible
even at rather large $Q^2$ (see also \cite{bar}). 

\begin{figure}[h]
\begin{center}
\leavevmode
\epsfig{file=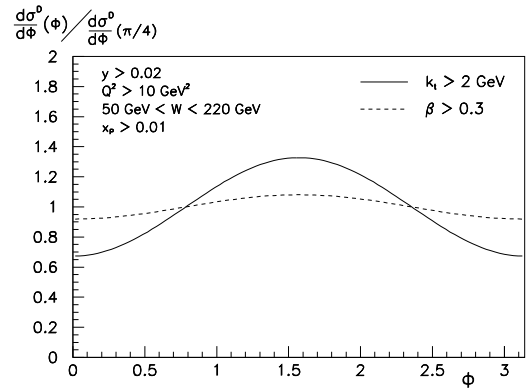,height=6cm}
\end{center}
\caption{Azimuthal distribution.}
\label{phifig}	
\end{figure}

Finally we examine the azimuthal distribution (Fig.\ref{phifig}). 
The azimuthal angle is defined in the transverse plane with respect
to the lepton. The interesting feature is the peak at $\pi/2$ which is
strongly enhanced when a hard final state is required, 
i.e. exclusive jets without a remnant (cut on $k_t$). 
But already the inclusive final state shows some oscillation (dashed line). 

\end{document}